\documentclass[journal]{vgtc}                     
\usepackage{hyphenat}

\onlineid{0}

\vgtccategory{Research}

\title{Exploring Object Recall in an Object-Rich AR Office Scene}

\author{%
  \authororcid{Sangita Kunapuli}{0009-0008-2745-1828},
  \authororcid{Shane Dirksen*}{0009-0001-1493-8169}, 
  \authororcid{Radha Kumaran*}{0000-0001-7161-3048}, 
  \authororcid{Yanxiu Jin}{0009-0008-4656-1555}, 
  \authororcid{You-Jin Kim}{0000-0003-0903-8999}, 
  \authororcid{Misha Sra}{0000-0001-8154-8518},\\
  \authororcid{Tobias Höllerer}{0000-0002-6240-0291}
}

\authorfooter{
  \item Dirksen and Kumaran contributed equally to this paper.
  \item
  	Sangita Kunapuli, Radha Kumaran, Shane Dirksen, Misha Sra and Tobias Höllerer are with the Department of Computer
Science, University of California, Santa Barbara. E-mail:
thollerer@ucsb.edu.
\item
  	Yanxiu Jin is with the Department of Electrical and Computer Engineering, University of California, Santa Barbara.

  \item You-Jin Kim is with the School of Computing, College of Engineering, University of Nebraska--Lincoln.
  \item This is a preprint version of this article. The final version of this paper can be found in IEEE TVCG (Vol. 32, 2026), presented at IEEE ISMAR 2026. For citation, please refer to the published version. This work was initially made available on the author's personal website [yujnkm.com] in September 2026, and was subsequently uploaded to arXiv for broader accessibility.
}

\abstract{%
  Augmented reality (AR) changes how users perceive and attend to their surroundings. Prior work suggests AR can reduce attention to the physical world, but it remains unclear whether this holds in environments dense with both virtual and physical objects, and which factors most affect object recall. This study examines recall in a mixed-reality office scene after an object-selection task, testing object-level factors (virtuality, size), context-level factors (semantic congruency, saliency), and user-level factors (cueing, dual tasking). Cueing informed participants of a later recall test, while dual tasking added cognitive load through a secondary audio task. Virtual objects were recalled more accurately than physical ones. Cueing produced a modest improvement in recall of present objects. Using playback software to reconstruct each participant’s view throughout the trials, we computed object-level saliency and found a weak positive correlation with recall, slightly stronger than the effects of virtuality, but not reaching statistical significance. These findings have implications for AR interface design and for building mixed-reality environments that better support cognition during real-world tasks.
}


\teaser{
  \centering
  \includegraphics[width=\linewidth]{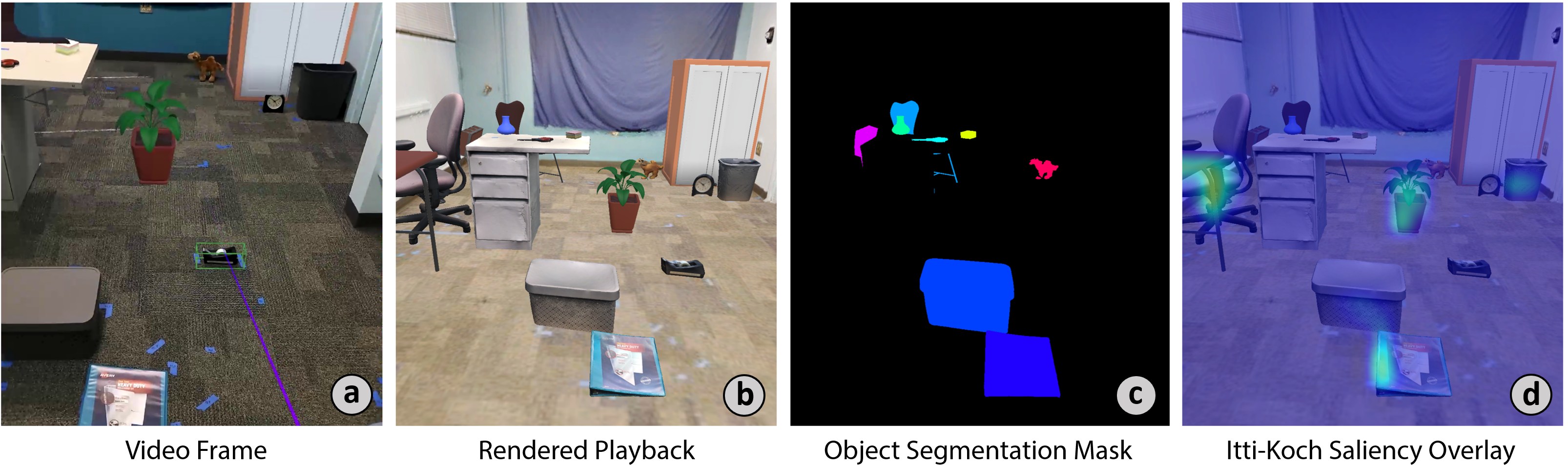}
  \caption{Overview of the data processing pipeline used to compute object-level saliency from participant viewpoints. (a) A captured video frame of the mixed-reality office environment, populated with both physical and virtual objects. (b) A rendered playback reconstructing the user's exact perspective during the trial. (c) Object segmentation masks used to isolate specific physical and virtual items within the field of view. (d) An Itti-Koch saliency map overlaid on the reconstructed scene enables us to compute the visual saliency of each object and assess its impact on user recall.}
  \label{fig:teaser}
}

\renewcommand{\manuscriptnotetxt}{}

\graphicspath{{figures/}{./}} 

\usepackage{booktabs}

\usepackage{mathptmx}                  

\begin{document}


\firstsection{Introduction}

\maketitle

Augmented reality (AR) enables the seamless integration of virtual content within real world environments, offering a wide range of applications, including remote work and productivity \cite{Li:2024:ARE,jeong2020using}, office environment simulations \cite{liu2025assessing}, contextual prototyping \cite{freitas2020systematic}, medical modeling \cite{Yeung:2021:VAR}, and advertising \cite{Du:2022:ARM}. These modern examples of AR use cases illustrate the versatility of the technology to be applied directly within a user's environment. AR headsets, such as the Magic Leap 2, use optical see-through technology, allowing users to perceive their physical environment through transparent waveguide displays, with virtual content layered on top and seemingly integrated. This implementation of AR maintains realistic spatial depth and lighting, allowing for a mixed reality experience comparable to viewing the real world~\cite{curtis2022unveiling}, but with virtual content added. It is an open question as to how AR users attend to this combination of virtual and physical objects, and the exact nature of this attention direction may have important consequences for AR applications. 

Several studies have found that AR users attend differently to virtual and physical objects (such as ~\cite{saliency2022, sutton2022look}). This could be problematic when there are important physical objects in the environment that get overlooked because attention is focused predominantly on virtual AR objects. If that were universally the case, AR designers should consider coping mechanisms, such as AR highlighting of mission-critical physical objects. 
But do those effects still occur when we have a world densely populated with virtual and physical objects, and what are other possible influence factors to object recall?
We designed a controlled study that sheds light on that question. We consider object recall (a user's report on having seen a particular object) as a conservative proxy for attention. In order to correctly recall an object (in the absence of guessing), the user must have attended to it at some time. But what factors influence object recall?

This work explores how object-level (virtuality, size), context-level (semantic congruency, saliency) and user-level (cueing, dual tasking) factors affect a user's ability to remember objects within a mixed reality office scene after completing an object selection task.


For the study in this paper, we designed different layouts of an object-rich mixed reality office scene to investigate the impact factors for object recall. After completing an object search task in the environment, participants complete the first ``surprise recall task." They then receive a verbal cue instructing them to form a mental model of their environment for the next memory recall task. This prompting allows us to study whether cueing improves participant ability to recall objects in a mixed reality scene, and how this effect depends on object-level (virtuality, size), context-level (semantic congruence, salience) and additional user-level (confidence, response bias, task prioritization) factors. To further probe attentional demands, we incorporate a secondary audio task in half of the trials as a source of additional cognitive load alongside the object search/selection task in our densely curated office scene. 

Our findings reveal that for our five modeled factors (virtuality, size, congruence, cueing, dual tasking) there are main effects of virtuality, driven by interactions of virtuality with size and congruence respectively: For large and medium-sized objects, virtual objects are recalled better than physical objects, and among semantically incongruent objects, virtual objects are also recalled better than physical objects. We thus see a confirmation of previous findings from the literature that virtual objects indeed appear to gather more attention than physical objects, under certain circumstances. 

We also report a small main effect of cue, with recall being better post-cue for present objects.


We assess recall sensitivity using signal detection theory and show that participant-reported recall significantly differs from chance guesses. With the effect sizes of our significant findings from the five modeled factors being quite small, we decided to explore additional impact factors and, using study playback software to reconstruct every participant viewing frame of the augmented scene throughout the trials, we computed object-level saliency and found a weak positive but statistically inconclusive correlation with recall.

Overall, cueing influences attention to objects present in the scene, interacts with decision strategies and confidence, and is partially guided by object semantics and participant priorities. From a holistic view, these results provide design guidance for AR systems and contribute to a deeper understanding of how attentional guidance, object level features, and intrinsic factors shape human cognitive processes and memory in mixed reality environments.



 \begin{figure*}[t]
\centering 
\includegraphics[width = 1.0\textwidth]{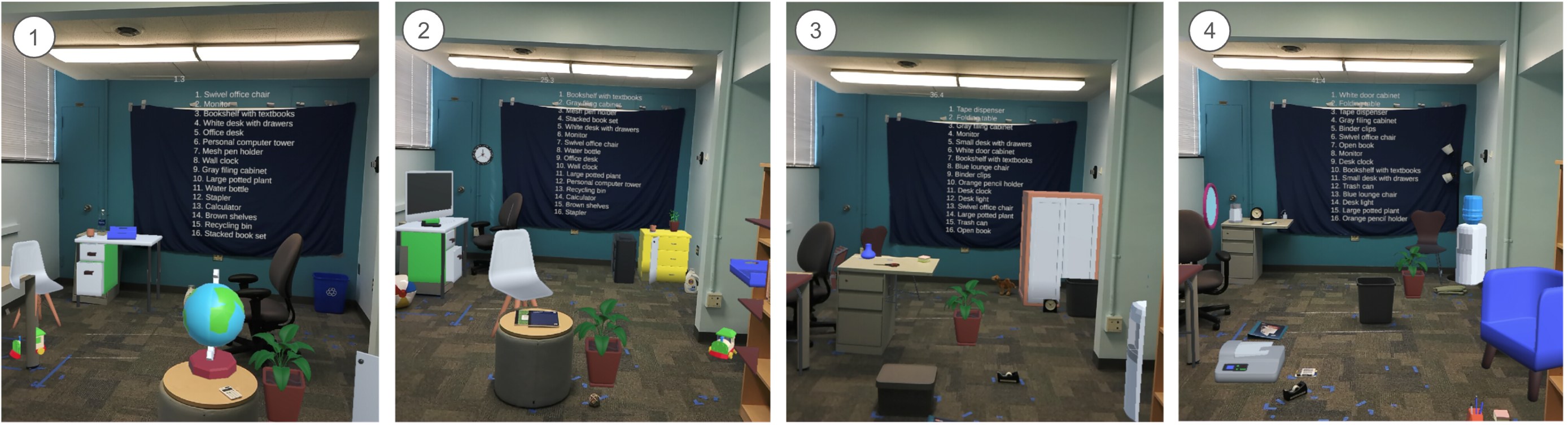}
 \caption{The four layouts consisting of different sets of objects. Layouts 1 and 2 were part of set A, while layouts 3 and 4 were part of set B. Each set consisted of an equal number of virtual and physical objects (16 each), equal proportions of objects by size, and 4 semantically incongruent objects in each set. Objects were placed at different locations in the room for each layout. Participants viewed, through the AR device, a list of objects they were supposed to find and select in sequential order.}
 \label{fig:fourlayouts}
 \end{figure*}

\section{Related Work}

\subsubsection*{Memory and Cognitive Load}

What people remember about an environment is shaped not only by their cognitive state, but also by properties of the objects themselves. 
One important object-level impact factor to memory is object-scene semantic congruency, which we use to operationalize {\em schema consistency}: whether an object fits expectations for a familiar environment~\cite{brewer1981role}. 
Prior work shows mixed effects of schema consistency on memory. Schemas can support memory for expected objects and also produce false memories for expected-but-absent objects \cite{brewer1981role,webb2020memory}, while unexpected objects can receive prioritized visual processing during encoding \cite{loftus1978cognitive,friedman1979framing} and can produce stronger recollection \cite{lampinen2001recollections} or recall \cite{pezdek1989memory}. Which mechanism dominates depends on methodological factors such as 
memory measure, and task context \cite{rojahn1992memory}. These schema effects also extend to immersive environments. In an HMD-based virtual office, schema-consistent objects were recognized more accurately and with higher confidence than schema-inconsistent objects \cite{mania2005effect}. Prior object-scene stimulus databases have measured this relation using participant ratings of semantic consistency \cite{ohlschlager2017scegram,andrade2024obscene}. A recent AR object-recall study similarly used rating-based thresholds to classify recall probes as congruent or incongruent while organizing recall objects by size categories such as small, medium, and large \cite{kumaran2025scene}.

Object size is another important property to control because prior work in scene memory has found that object size significantly predicts whether objects are recalled from scenes \cite{hall2025object}, and size can also influence attentional capture during visual search \cite{proulx2010size}. Using discrete size classes is psychologically grounded: people make highly consistent judgments about relative object size, and these judgments scale with the logarithm of objects' real-world size \cite{konkle2011canonical}. In AR, virtuality is another relevant dimension: prior AR search work found lower recall for physical than virtual objects, suggesting that memory can differ across physical and virtual content \cite{kim2022investigating}. 

Hence, we include a semantically incongruent subset of objects, identified through office-congruency ratings, to investigate how schema-based expectations affect memory in AR. We also categorize and balance our object sets by size bin and virtuality to control object-level factors that may affect recall.

Cognitive load has also been shown to have a negative effect on memory performance. Researchers have investigated how multitasking affects memory recall in an investigative interview task and found that higher cognitive load diminished participant ability to recall key information \cite{Hanway:2021:CLI}. The results from this study suggest that in scenarios with increased cognitive load and multitasking, limited cognitive resources can constrain memory performance. More broadly, attention during encoding plays an important role in later memory: dividing attention with a concurrent auditory task can impair recollective recognition \cite{gardiner1990attention}, while intentional encoding instructions can improve recognition of objects in real-world rooms relative to incidental exposure \cite{pezdek1989memory}. These effects can be further shaped by prioritization and motivation, particularly when information is tied to a task demand or reward \cite{myers2017prioritizing, grogan2022motivation}. In this study, we investigate how memory performance may be influenced when verbally cued (prioritized) and under various cognitive demands.

\subsubsection*{Attention in Augmented Reality}

AR can support a range of tasks with spatially relevant annotations that present users with necessary information or cues. Prior work has examined such interfaces for applications including navigation and wayfinding~\cite{kelley20242d}, industrial workflows~\cite{malta2023augmented} and education~\cite{ibrahim2018arbis}. Introducing virtual annotations into a user's view of the real world may change how attention is directed, and potentially increase cognitive load compared to non-AR problem-solving, even if AR does improve performance in the task at hand~\cite{buchner2022impact, alessa2023neurophysiological}. Prior work has also demonstrated varying recall of physical and virtual environment objects based on task focus~\cite{Kim:2025:AGO, Kumaran:2023:NAR}, suggesting that object-level features that impact visual salience could alter attention direction in AR.

A driving simulator study looked at the potential of cues within an AR interface to shift driver attention to roadside hazards. The researchers investigated impacts on driving ability and found that it showed an increase in response time and limited interference, indicating that with proper design paradigms, directed attention within AR can prove useful in a high demand situation \cite{rusch2013directing}. In other work exploring the potential of AR for object recall~\cite{munoz2018augmented}, touring an augmented environment helped participants to better remember the location of virtual objects added to the real scene compared to looking at photographs of the environment. 

These studies highlight the importance of studying factors impacting attention in AR systems. They indicate that memory and perception are dependent on AR interface design, semantics of the environment, and object features. There is limited understanding of how attention is directed intentionally or unintentionally during AR task loads. Our work adds to this body of work by focusing on cueing, object factors, and multitasking to evaluate memory and attention in a controlled attention-demanding real-world mixed reality task scenario.

\begin{figure*}[t]
  \centering
  \begin{minipage}[b]{0.48\textwidth}
    \centering
    \includegraphics[width=\linewidth]{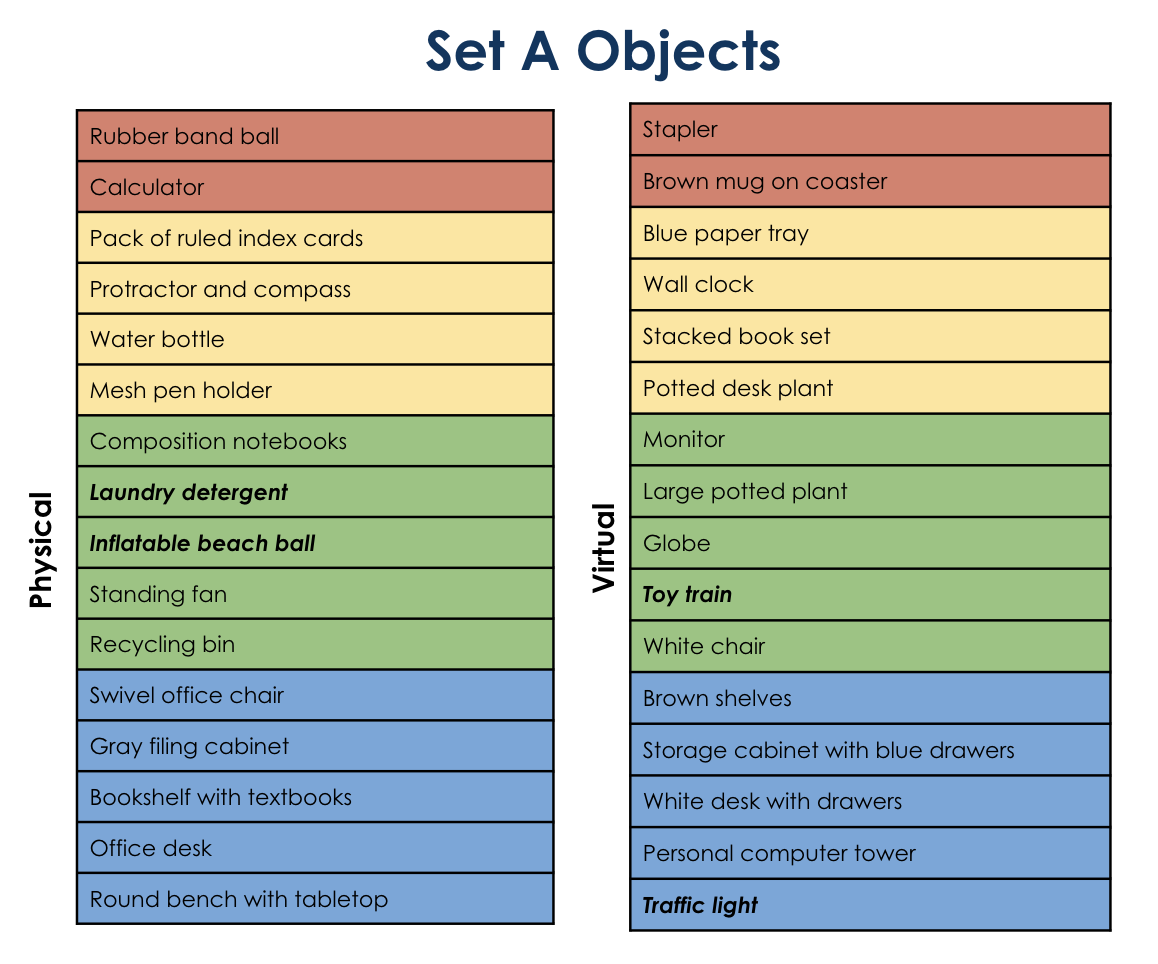}
 
  \end{minipage}\hfill
  \begin{minipage}[b]{0.48\textwidth}
    \centering
    \includegraphics[width=\linewidth]{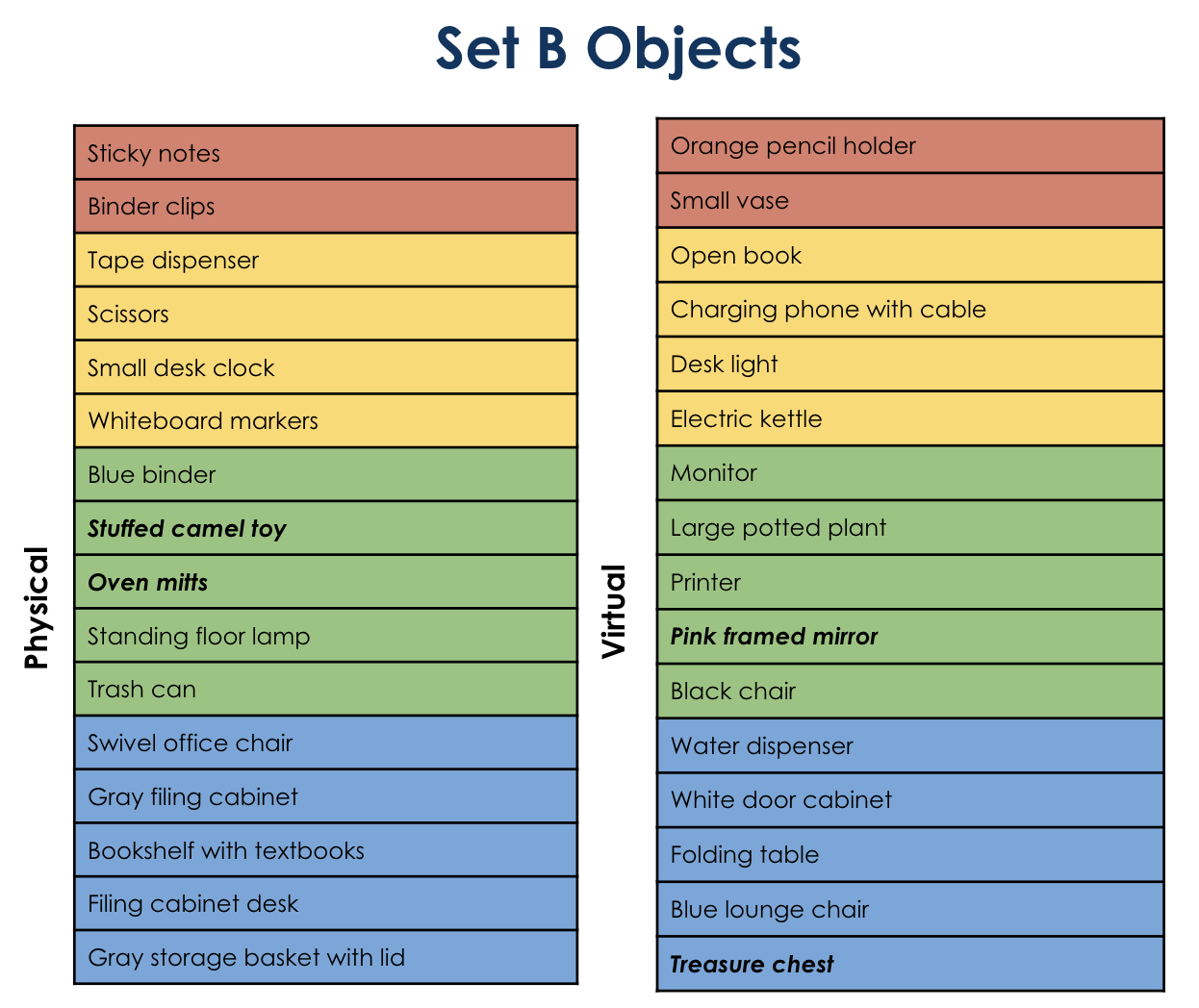}

  \end{minipage}
  \caption{The exact objects that show up in the environment and are part of each set. Set A corresponds with layout 1 and 2, while set B corresponds with layout 3 and 4. The list is ordered from smallest to largest size and split evenly by virtuality. The size categorizations range from very small (red), small (yellow), medium (green), and large (blue). The objects listed in bold are the incongruent ones of each set. There is a 1:1 relationship between objects in set A and B to minimize bias.}
      \label{fig:setBobj}
\end{figure*}

\subsubsection*{Saliency in Augmented Reality}

Visual saliency is the subjective perceptual quality that makes certain locations or objects stand out from their surroundings and immediately grab our attention \cite{veas2011directing}. Saliency is commonly understood to have two attentional selection factors: bottom-up, which is stimulus-driven and emerges automatically from low-level feature differences in dimensions such as color, orientation and motion; and top-down, which reflects cognitive processes such as volitional, goal-directed control and selection driven by prior experience. Bottom-up saliency has been widely exploited in AR across several research directions.

One line of work explores saliency modulation in AR to guide user attention by enhancing target saliency \cite{ahn2018real}, suppressing distractors \cite{koshi2019augmented,lukashova2021saliency}, or combining both \cite{veas2011directing,sutton2022look}. Early work introduced Saliency Modulation Technique (SMT) in live videos, showing accelerated fixation on target regions and improved object recall \cite{veas2011directing}. Subsequent studies extended this concept to mobile AR applications such as improving AR imagery visibility \cite{ahn2018real} and guiding attention in educational contexts \cite{mcnamara2011enhancing}. Sutton et al. \cite{sutton2022look} later implemented the first saliency modulation technique for optical see-through (OST) HMDs, demonstrating increased fixation likelihood and reduced time to first fixation, while other work in HMDs suppressed distractors to improve concentration \cite{koshi2019augmented} and visual search performance \cite{lukashova2021saliency}. Another line of work studies how the saliency of visual cues themselves can be designed to attract user attention in AR. Prior works have explored effective cues such as hue-based dichoptic cues \cite{erickson2022analysis} and enhanced contour or dynamic saliency which have been shown to improve visual search efficiency and reduce inattentional blindness in AR-HUD navigation scenarios \cite{zhu2025visual}. 

Beyond direct attention guidance, researchers have leveraged saliency computation to estimate a user's attention distribution and determine the optimal spatial positioning of virtual information, balancing information visibility while minimizing occlusion with visually important areas of the real-world scene \cite{chang2016automatic, rakholia2018place}. Recent works have also investigated saliency prediction under AR visual confusion, showing that attention distribution is jointly influenced by factors regarding background scenes, augmented content, their mixing ratio, and temporal dynamics \cite{xie2025visual}. While these efforts have advanced the understanding of bottom-up saliency for attention guidance in AR, little is known about how object salience influences memory. In our study, we investigate the relationship between object salience and corresponding recall performance, providing insights for AR interface design that goes beyond directing attention to supporting users’ memory of their surrounding objects.

\subsubsection*{Dual-Tasking Attention Demands}

The dual-task paradigm examines how performing two tasks concurrently strains limited attentional resources, typically producing slower reaction times and reduced accuracy \cite{pashler1989chronometric}. While resource theories suggest that interference may be minimized when tasks draw from different modalities such as vision and audition \cite{wickens1984processing}, evidence shows that cross-modal demands still impair performance. For example, simultaneous auditory and visual tasks reduce efficiency, indicating that attentional capacity is not fully independent across modalities \cite{spence1997audiovisual}. Real-world findings, such as the detrimental effect of cell phone conversations on driving, confirm that auditory distraction disrupts selective visual attention, degrading task performance \cite{strayer2001driven}.

In AR, these dual-task limitations are particularly relevant, as headsets often integrate multimodal input. Visual demands imposed by AR displays compete with concurrent auditory streams, and binocular AR conditions have been shown to reduce manual task accuracy \cite{condino2019perceptual}. AR performance under dual-task conditions can be constrained by both cross-modal interference and the intrinsic demands of  the device. In this study, we purposefully employ an auditory call sign task to create an attention-demanding environment in which to explore influence factors on AR object recall.  

\section{Experiment}

\subsection{Study Design}

Participants put on an AR headset in an object-rich (cluttered) office environment and completed a sequence of eight 45-second trials during which they were tasked to spot and select objects from a list displayed on one of the office walls. 

In each trial, participants completed a standing selection task from a fixed spot in the office. They were presented with a virtual list of 16 objects and must sequentially find and select, by pointing and clicking with the controller, as many objects as possible in the mixed reality office scene within 45 seconds. There were 8 trials total. Participants completed a surprise memory recall task after the 4th trial, were subsequently instructed to create a mental model of their environment through a verbal cue, and finally completed the cued memory recall task after the 8th trial.

\subsubsection{Variables}
\label{sec:variables} 

We examined a number of independent variables in this study, including object-, user- and context-level factors.

Our object-level factors of interest were the \textit{virtuality} of an object [objects were either physical or virtual] and the \textit{size} of the object [large, medium, small, very small]. Of the 32 objects in the scene in each trial, 16 were physical, and participants were tested for recall of 8 physical and 8 virtual objects at the end of their trials. The remaining physical and virtual objects were included in the primary selection task, to ensure that participants were not biased to recall of certain objects due to the selection task alone. The objects were also distributed across a range of sizes, though we could not achieve a perfectly balanced design with regards to the size variable, due to the physical constraints of the office-themed space.

The user-level factors we were interested in were cueing (user alerted to the recall task) and dual task (additional audio task in half the trials). We define cueing as a verbal cue directing participants to remember objects in the scene. All participants complete the memory recall task after the first 4 trials without prior knowledge (no cue condition). Participants are then verbally cued (cue condition) to remember all objects in the scene, and complete another 4 trials in the environment. We use a secondary audio response task (participant pressing a controller button when they hear a certain radio alphabet letter) that goes on in conjunction with the selection task as a means of increasing cognitive load. The cue condition is a within subjects variable, so all participants experience the no cue and cued conditions (in that order, since counterbalancing is impossible in this case). Participants experience the audio task for all trials of the first or second block of 4 trials, counterbalanced across participants.

We also studied context-level factors such as semantic \textit{congruency} and saliency. In each set of 32 objects, four of the objects were semantically incongruent (2 physical and 2 virtual). We chose to limit the number of incongruent objects since including a larger number of incongruent objects (half the objects, for example) would impact the semantics of the entire scene and hence potentially change the perception of congruency. We modeled object saliency after the experiment was completed, using motion data collected during the trials and replayed in our custom playback software. This allowed us to compute more precise saliency values for each participants' perceptions of the objects based on their head movement and viewing angles.

\begin{figure}[t]
    \centering
    \includegraphics[width=1.0\linewidth]{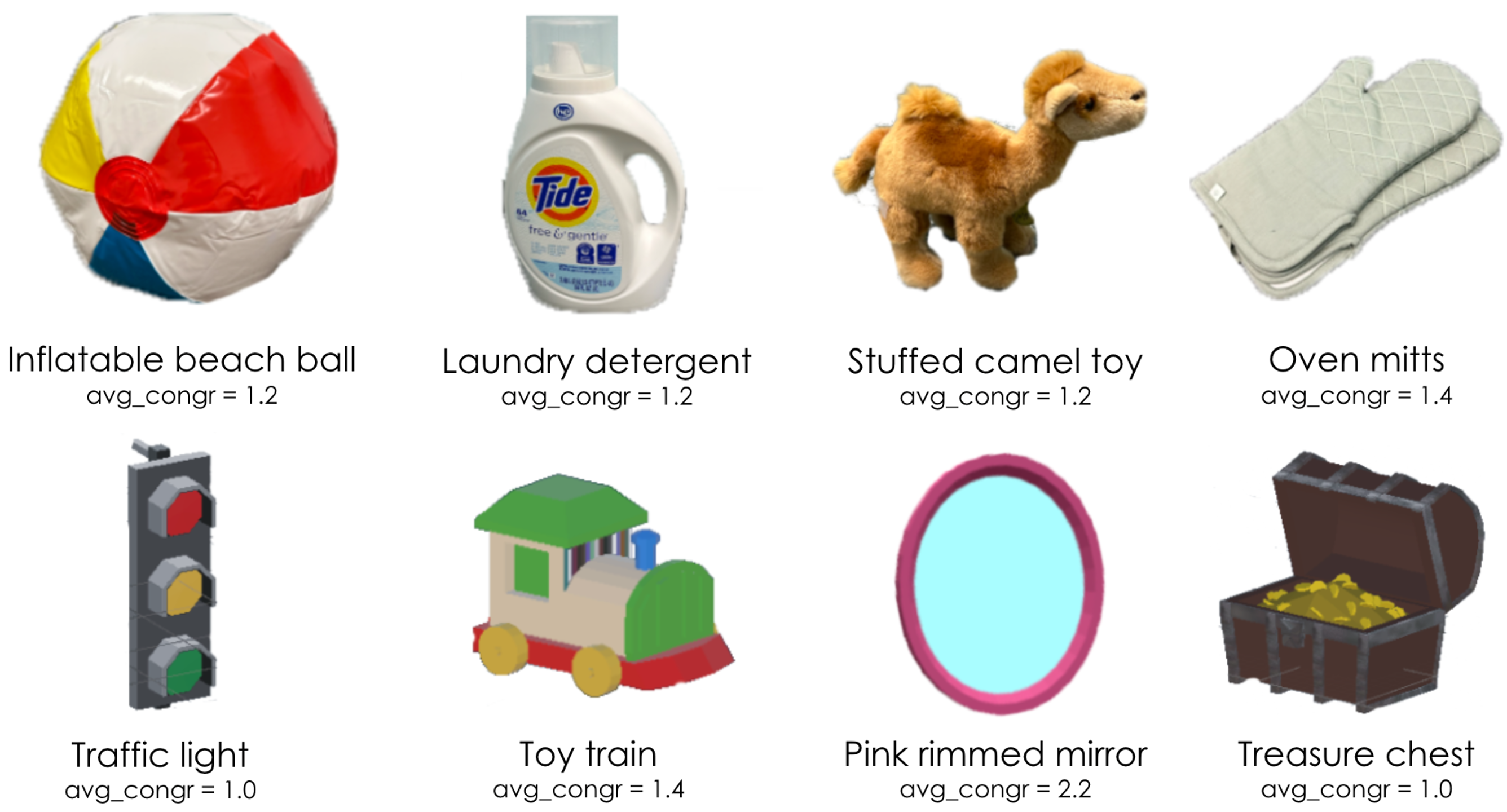}
    \caption{Semantically incongruent objects. Each object received an average congruency ranking (\texorpdfstring{avg\_congr}{avg congr}) based on our object congruency survey. Objects with average congruency ratings less than 2.5 were considered incongruent.}
    \label{fig:incongruent}
    \vspace*{-3ex}
\end{figure}

\subsubsection{Objects} 
We use two different sets of objects in this study, as shown in \autoref{fig:setBobj}. A set of objects consists of 32 objects total that appear in the scene: 16 physical and 16 virtual. All of the objects in the room during the experiment are considered part of the set, excluding room-frame objects such as light switches, walls, windows, and doors. The sets of objects are balanced in terms of virtuality (physical vs. virtual) and size (very small, small, medium, large) across virtuality. Objects were categorized into the four size bins based on overall physical dimensions to ensure balanced representation of object scale across physical and virtual conditions. Based on the longest object dimension, the resulting categories had median sizes of 13.9 cm (very small), 28.0 cm (small), 46.9 cm (medium), and 93.7 cm (large), with ranges of 8.8–19.1 cm, 16.0–57.0 cm, 27.5–168.9 cm, and 50.2–203.2 cm, respectively. A set of objects contains 4 incongruent objects, 2 physical and 2 virtual. In this context, congruency determines whether an object is semantically congruent with an office space. For example, a beach ball would be considered incongruent in an office space, while a printer would be considered congruent. In order to determine our set of incongruent objects, we conducted an object congruency survey with 5 participants and had each participant rank how congruent they thought each object was in an office scene and what they would call the object (in order to arrive at a generally accepted moniker to refer to objects on the search list and recall task). The results from this survey are shown in \autoref{fig:incongruent}, and we use a threshold of $avg\_congr < 2.5$ to determine incongruity. We employ GPT4o, prompted with images of the objects, and the object congruency survey results to determine the most universally understood descriptor names for the objects, and we use these consistently throughout the tasks.

\subsubsection{Selection Task} 
Participants are given 45 seconds to sequentially find and select up to 16 objects from a list provided to them during each trial. Participants stand in a doorway looking into the room and are informed that objects in the scene can show up anywhere within the room, covering a hemisphere in front of them, including on surfaces and the floor. They must search around the room, pointing at and selecting both physical and virtual objects with the controller. During the selection task, participants receive visual feedback of a bounding box around the object appearing when the controller ray intersects with it, which turns green on correct selection. The object name on the list is also crossed off immediately. Of the 32 objects that are part of the scene, 16 are part of the selection task.

\subsubsection{Dual Audio Task} 
For the audio task, participants were expected to listen to a randomized stream of 5 words from the NATO phonetic alphabet (alpha, bravo, charlie, delta, echo) that played in the headset at intervals of 3 seconds. Whenever they heard the word \textit{charlie}, participants were supposed to respond by pressing the menu button on top of the Magic Leap 2 controller. The audio task was included in half of each participants' set of 8 trials to apply additional cognitive load. Participants were informed prior to each trial if there would be an audio task.

\subsubsection{Memory Recall Task} 
Participants are asked about 32 objects on the memory recall task. 16 of these objects were present in the scene (but not part of the selection task), and 16 objects were not part of the scene (absent). For each object on the task, participants were asked if the object was present or absent, physical or virtual (or not present if they selected absent), and their associated self-confidence rating (on a scale of 1-10) for their answers. After participants completed the memory recall task, they were shown their score on the task to implicitly motivate prioritization of the task.

\begin{figure*}[t]
\centering 
\includegraphics[width = 1.0\textwidth]{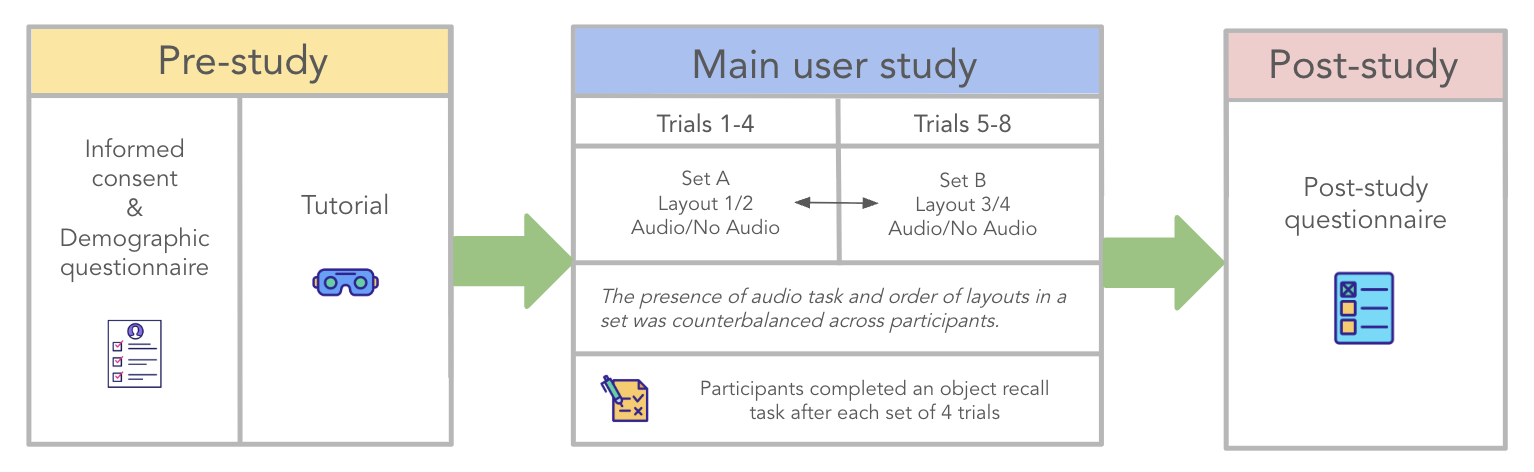}
 \caption{Schematic representation of the study flow. The user study took place in three stages: pre-study preparation, the main study trials, and post-study questionnaire. In the main user study, participants either experienced Set A or Set B for their first set of trials and their corresponding layouts. The surprise object recall assessment took place after the 4th trial. }
 \label{fig:studyschematic}
 \end{figure*}

\subsection{Methodology}
 \autoref{fig:studyschematic} shows the overall study procedure.  
 Participants first signed the informed consent form and completed the pre-study demographic questionnaire to gather data on their experience with AR/VR and sense of direction (SBSOD) \cite{hegarty2002development}. They were then provided with slides containing instructions for the selection task and information on how to operate the Magic Leap 2 device and controller. After fitting the headset comfortably, participants did a short eye tracking calibration and practiced selection and the audio task with the controller simultaneously. Participants were shown multi-colored 3D cubes to practice selection on, so that they were not exposed to office objects prior to the trials. Each trial was completed standing up within a designated region at the entrance of the room. Participants were instructed to view the mixed reality office environment as a ``display" and told that objects could show up anywhere inside the room. They were not allowed to walk into the environment to minimize any extra load of walking and standardize the experience between participants. Before completion of the trials, participants were not informed that they would be completing a memory recall assessment. 

The first 4 trials consisted of the same set of objects but in two different layouts, as shown in \autoref{fig:fourlayouts}. After the 2nd trial, the layout changed: the same objects show up in different locations. Object placement was randomized to the extent possible, given the physical constraints of the environment. Large and heavy objects, such as the bookshelf, could not easily be repositioned and stayed fixed. To balance spatial exposure across layouts, objects that were closer and central to the participant in one layout were generally placed farther away or at the perimeter in the other layout, and vice versa. The incongruent objects also moved around the room to ensure they were as visible as the congruent objects. Smaller objects, which are inherently more difficult to identify when further away, were placed in visible locations relative to their size while maintaining layout distribution.  Participants completed each layout twice with 2 different lists to mitigate bias from list difficulty and to maximize visibility of objects in the scene. The order of layouts and object lists was counterbalanced across participants using a Latin Square design. This was to ensure that each layout and list appeared equally often while balancing potential order and learning effect bias across participants. After the 4th trial, the set of objects changed while participants completed their initial surprise memory recall assessment based on their first set. They were asked about objects they did not directly interact with through selection.

The verbal cue was administered after the 4th trial once participants had completed the first memory recall assessment. Participants were shown their performance on the first memory recall assessment to incentivize performance for the second set of trials. The verbal cue is: \textit{``In this next set of trials, while completing the selection task to the best of your ability within the time constraint, please focus on creating a mental model of the objects in the room."}

Following completion of all trials and recall assessments, participants completed a post-study questionnaire related to their experience. As part of this questionnaire, participants ranked their relative prioritization for each of the three concurrent tasks: object selection, auditory task, and memorizing the room for later recall. These responses were used to assess participants' perceived task prioritization during the experiment.

\subsection{Participants}
The study involved 32 participants aged 18-30 years, consisting of 18 female and 14 male. We conducted an a-priori power analysis using G*Power for a repeated measures within subject ANOVA\footnote{We based the power analysis on ANOVA parameterization even though we eventually opted for binomial GLMMs as our primary analysis tool.} ($f=0.27$, $\alpha=0.05$, power $=0.95$, 4 repeated measures) which estimated a sample size of 31 participants. Therefore, we recruited 32 participants to achieve complete counterbalancing of experimental conditions across participants and satisfy power. Of the 32 participants, 29 were right-handed, and 3 were left-handed. 7 participants wore contacts for the study to have corrected-to-normal vision. Participants had varying familiarity with VR/AR: over half of the subjects had not used AR previously (56.3\%). In comparison, 87.5\% of participants had used virtual reality one or more times. Participants were compensated at \$15 per hour. All procedures were approved by UCSB's Human Subjects Committee (protocol \#53-25-0164).

\subsection{Analysis}

We measured \textit{recall accuracy} as a binomial metric, with either correct or incorrect responses to the question about the object being present in the scene. We also recorded \textit{virtuality accuracy} as a binomial metric of whether they correctly remembered the virtuality of the objects they indicated were present. Lastly, participants reported a \textit{confidence score} for the responses to each object on a 10-point Likert scale ("Not confident" to "Extremely confident"). 

Recall accuracy was modeled as a GLMM with a binomial distribution, with participant ID as a random effect and combinations of our independent variables (object virtuality, object size, object congruence, cueing, dual task) as fixed effects. Results are reported with Type-II ANOVAs due to the unbalanced dataset (far more congruent objects than incongruent ones) and post-hoc pairwise comparisons were performed with Bonferroni corrections applied per cluster of comparisons. The GLMM was implemented using the \emph{lme4} package in R, and post-hoc comparisons were performed using the \emph{emmeans} package. Correlation analyses were performed using Spearman's Rank Correlation coefficient.

\section{Results}

In accordance with our main research goals, we divide our results into three main areas: impact of the {\em object-level factors} (virtuality, size) on recall, {\em user-level factors} (cueing, dual tasking), and finally, of the {\em context-level factors} (congruency, saliency) on recall performance. Our study was conducted initially as an analysis of the main factors: virtuality, size, congruency, cueing, and dual tasking to serve as the primary variables of the study. The subsequent saliency analysis was conducted as an exploratory investigation to better understand potential sources of variance not captured by the modeled factors, particularly given the inherently multi-factorial nature of object recall in mixed-reality environments. 

We conclude the result section with a table of the factors that had the largest correlation with object recall.

\subsection{Signal Detection}

Because we set up the task scenario as a purposefully challenging object recall task in an office environment densely populated with physical and virtual objects, and we thus expected merely modest recall performance, we first confirmed that the distribution of responses we observed was significantly different from chance, and that there indeed is a meaningful recall signal for us to model and analyze. Overall recall accuracy amounted to 67.63\% across all recall events. 

We assessed recall sensitivity using signal detection theory\cite{stanislaw1999calculation}, treating present objects as signal and absent objects as noise; chance performance corresponds to d' = 0, so we tested whether participant-level d' values were significantly greater than zero. A one-sample Wilcoxon signed-rank test showed that d' values were significantly greater than zero (V = 528, p $<$ .001, one-tailed), indicating above-chance recall sensitivity.
This analysis partitions the variance in recall responses into a signal-driven component, bounded by participants' demonstrated discriminability, and a 
component of conditional response uncertainty (e.g. participant guesswork).
Because recall was a single-object present/absent decision, with the optimal cutoff sitting halfway (d'/2) between the two distributions, the estimated signal-driven accuracy of participant recall under an unbiased criterion is $\Theta(d'/2) = \Theta(0.49) \approx 0.689 = 68.9\% (d' = 0.98)$.

We tested several machine-learning methods on our data to create a predictive classifier for recall, each combining every measured predictor (all variables from Section \ref{sec:variables}). Our best model reaches 69.82\% prediction rate of actual human recall response.

\subsection{Object-Level Factors}

\begin{figure}[t]
    \centering
    \includegraphics[width=0.9\linewidth]{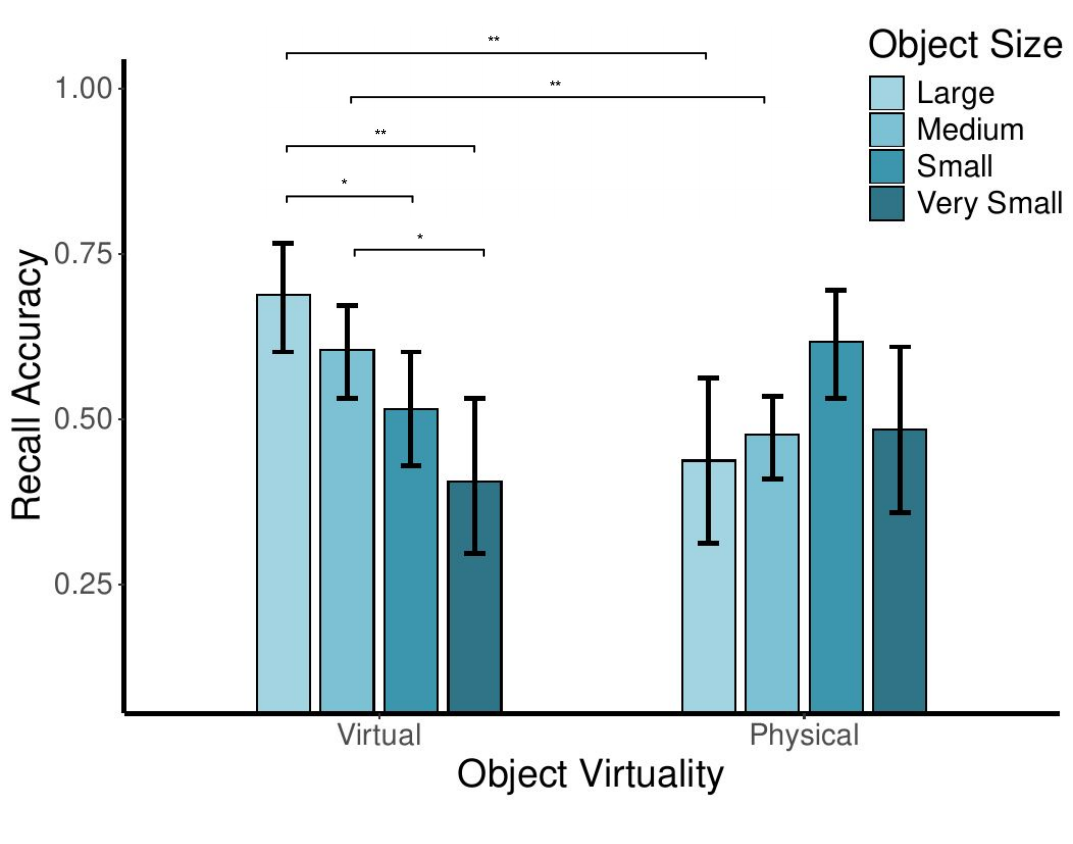}
    \caption{Recall accuracy for present objects as a function of object virtuality and size. For virtual objects, large objects were significantly better recalled than small and very small objects, and medium-sized objects were better recalled than very small objects. For large and medium sized objects, virtual objects were better recalled than physical objects. Error bars represent 95\% confidence intervals.}
    \label{fig:size-virtuality}
    \vspace*{-3ex}
\end{figure}

We analyzed the impact of the object-level factors using a binomial GLMM with object virtuality [physical, virtual] and size [large, medium, small, very small] as fixed effects and participant ID as a random effect. Object was not included as a random effect since it is confounded with the object size/virtuality manipulations examined in this model. Results are visualized in \autoref{fig:size-virtuality}. There was a main effect of virtuality ($\chi_{}^{\mathrm{2}}$(1) = 3.98, $p$ = 0.046, $\eta^2_{p}$ = 0.003) and an interaction between virtuality and size ($\chi_{}^{\mathrm{2}}$(3) = 18.06, $p$ $<$ 0.001, $\eta^2_{p}$ = 0.013), but no main effect of size ($\chi_{}^{\mathrm{2}}$(3) = 7.244, $p$ = 0.065, $\eta^2_{p}$ = 0.005). The conditional $R^2$ was 0.075 and the marginal $R^2$ was 0.037.

To break down this interaction, we examined size at each level of virtuality. Among the virtual objects, recall was significantly higher for large objects compared to small [$OR$  = 2.114, $SE$ = 0.558, $p$ = 0.028] and very small [$OR$  = 3.338, $SE$ = 1.079, $p$ = 0.001] objects. Recall was also higher for medium objects compared to very small objects [$OR$  = 2.29, $SE$ = 0.685, $p$ = 0.034]. None of the other comparisons at the virtual or physical levels were significant (all $OR$ $<$ 1.747, all $SE$ $<$ 0.549, all $p$ $>$ 0.051).

We also examined virtuality at each level of size, and found that virtual objects were better recalled than physical objects for both large [$OR$  = 0.342, $SE$ = 0.11, $p$ = 0.001] and medium sized objects [$OR$  = 0.586, $SE$ = 0.115, $p$ = 0.007], though there were no significant differences for small and very small objects (all $OR$ $>$ 1.387, all $SE$ $>$ 0.396, all $p$ $>$ 0.096).

\subsection{User-level Factors}

\subsubsection{Cueing and Audio Task Effects on Recall}

\begin{figure}[t]
    \vspace{0.1in}
    \centering
    \includegraphics[width=\linewidth]{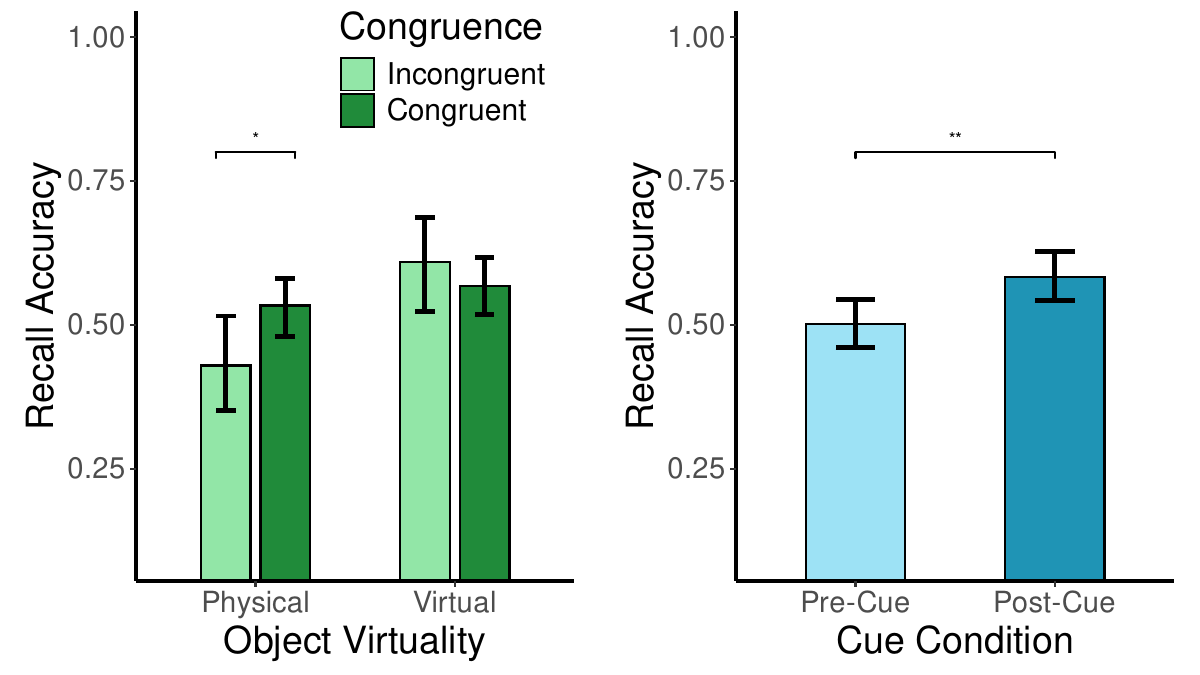}
    \caption{(Left) Recall accuracy for present objects as a function of object virtuality and object congruence. For physical objects, congruent objects were better recalled than incongruent objects. (Right) Recall accuracy as a function of cueing condition. Recall was significantly higher post-cue compared to pre-cue. Error bars represent 95\% confidence intervals.}
    \label{fig:cue}
    \vspace*{-3ex}
\end{figure}

We analyzed the impact of the user-level factors using a binomial GLMM with cueing [pre-cue, post-cue] and dual task [present, absent] as fixed effects, and participant ID and object as random effects. Including the object as a random effect improved the model compared to a model with only participant as random effect ($\Delta$AIC = 38, $\Delta$BIC = 33). Results are visualized in \autoref{fig:cue} on the right. There was a main effect of cue condition ($\chi_{}^{\mathrm{2}}$(1) = 7.70, $p$ = 0.005, $\eta^2_{p}$ = 0.006), with significantly higher recall post-cue compared to pre-cue [$\beta$  = 0.373, $SE$ = 0.133, $p$ = 0.005]. There was no main effect of audio task ($\chi_{}^{\mathrm{2}}$(1) = 0.88, $p$ = 0.347, $\eta^2_{p}$ = 0.001) or interaction ($\chi_{}^{\mathrm{2}}$(1) = 2.924, $p$ = 0.087, $\eta^2_{p}$ = 0.002). The conditional $R^2$ was 0.148, and the marginal $R^2$ was 0.017.

\subsubsection{Confidence} 

\begin{figure}[t]
    \centering
    \includegraphics[width=1\linewidth]{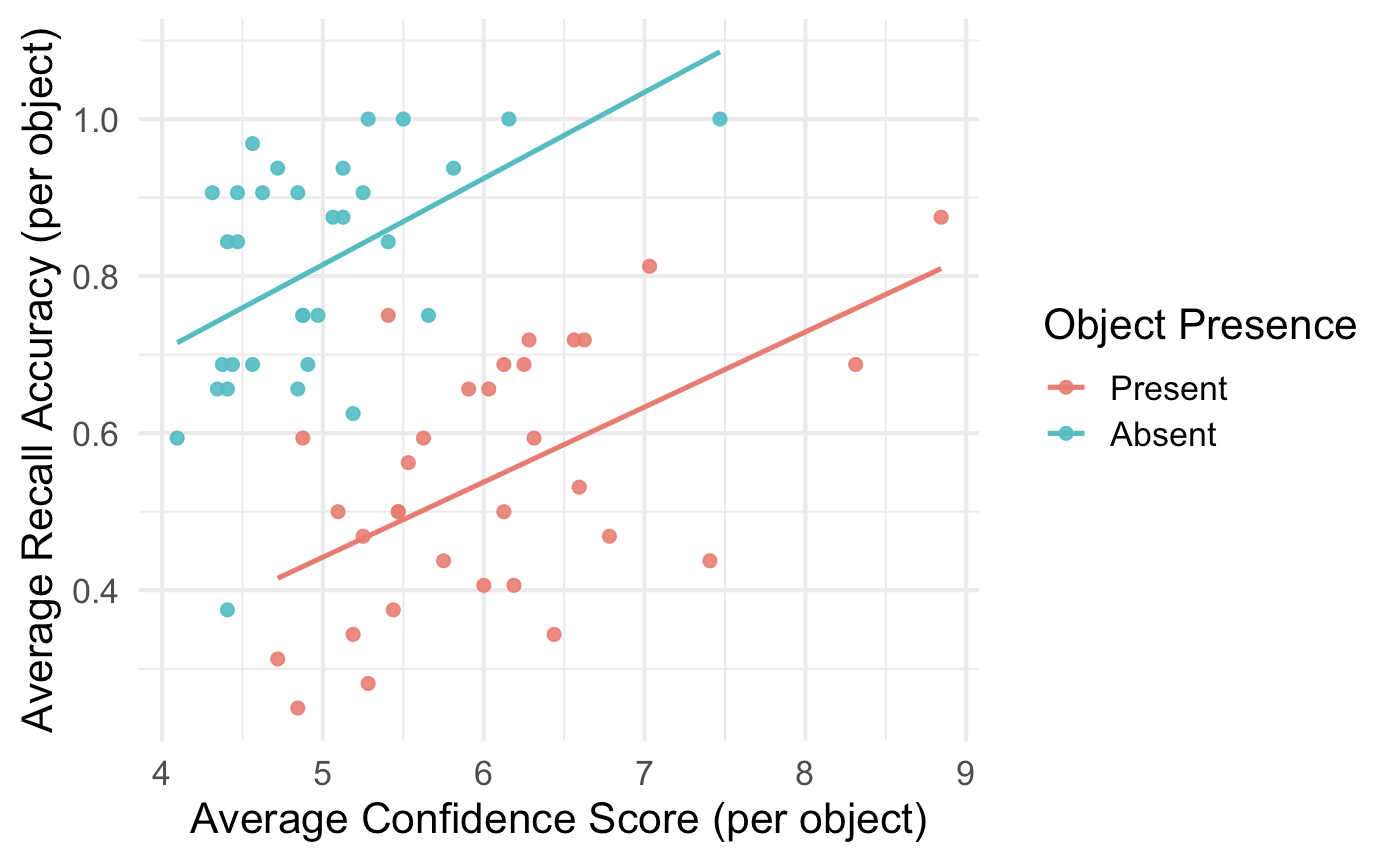}
    \caption{Recall accuracy as a function of participants' average confidence score, split by object presence. Each point represents an object on the memory recall assessment. Both present and absent objects have a positive correlation between confidence and recall accuracy, with absent object confidence tending to be lower, and present object accuracy tending to be lower.}
    \label{fig:presence-scatter}
    \vspace*{-3ex}
\end{figure}

To further investigate recall performance, we examine confidence scores to understand participant decision-making. Our analysis of confidence scores, as shown in \autoref{fig:presence-scatter}, reveals that participants were not simply guessing on the recall assessment, as the average confidence had a positive correlation with recall accuracy for both present $(r = 0.55, p = 0.00103, n = 32)$ and absent $(r = 0.49, p = 0.00435, n = 32)$ objects. We see that higher confidence tracks recall accuracy, with present objects having generally higher confidence and lower accuracy and absent objects having lower confidence but higher accuracy comparatively. Participants had higher confidence on incongruent absent objects, but lower recall accuracy compared to congruent absent objects. Participants had significantly different levels (p $<$ 0.0001) of confidence for incongruent and congruent objects and across object presence. Despite higher confidence, participants are influenced by congruence with the scene, and among incongruent objects, participants were more successful at identifying absent objects than present objects. We see specific examples of this phenomenon based on our results of the objects with the highest recall performance shown in \autoref{fig:saliency-recall}. The highest performers included the treasure chest, traffic light, toy train (present) and tennis racket, acoustic guitar, and princess themed music box (absent), all of which we consider incongruent with the office space.

\subsection{Context-level Factors}
\subsubsection{Congruency}

We analyzed the impact of the context-level factors using a binomial GLMM with congruence [congruent, incongruent] and object virtuality [physical, virtual]  as fixed effects and participant ID as a random effect. Object was not included as a random effect since it is confounded with the object congruence/virtuality manipulations examined in this model. Results are visualized in \autoref{fig:cue} on the left. There was a main effect of virtuality ($\chi_{}^{\mathrm{2}}$(1) = 5.151, $p$ = 0.023, $\eta^2_{p}$ = 0.004) and an interaction between virtuality and congruence ($\chi_{}^{\mathrm{2}}$(1) = 4.195, $p$ = 0.04, $\eta^2_{p}$ = 0.003), but no main effect of congruence ($\chi_{}^{\mathrm{2}}$(1) = 0.774, $p$ = 0.378, $\eta^2_{p}$ = 0.001). The conditional $R^2$ was 0.049, and the marginal $R^2$ was 0.012.

To break down this interaction, we examined congruence at each level of virtuality. Among the physical objects, recall was significantly higher for congruent objects compared to incongruent [$OR$  = 0.649, $SE$ = 0.136, $p$ = 0.037] but this was not true for virtual objects [$OR$  = 1.194, $SE$ = 0.252, $p$ = 0.402] objects. 

We also examined virtuality at each level of congruence, and found that virtual objects were better recalled than physical for incongruent objects [$OR$  = 0.472, $SE$ = 0.122, $p$ = 0.004] though there was no significant difference for congruent objects [$OR$  = 0.868, $SE$ = 0.128, $p$ = 0.338].


\subsubsection{Computational Saliency}

To generate accurate saliency metrics per participant, the virtual environment was reconstructed by first capturing the physical room using a Matterport 3D scanner to produce a geometrically accurate mesh of the space. Individual objects placed throughout the room were separately scanned using 3D Gaussian splatting and converted to meshes, then positioned within the Unity scene to match their real-world locations. During each trial, the headset recorded per-frame head position and orientation for each participant. To generate frame-level visual data for offline analysis, we replayed each participant's recorded head trajectory through the reconstructed scene, interpolating between consecutive frames to reproduce their exact viewpoint. At each frame, two synchronized renders were captured at 1920×1080: an RGB image from the participant's perspective and a corresponding object segmentation mask in which each of the 16 tracked recall objects (per layout) was rendered in a unique color (generated via evenly spaced HSV hues) while all other geometry was rendered in black. These paired RGB and mask frames were exported for each of the 256 trials across 32 participants.

Each frame was then processed through five computational saliency models spanning classical and deep learning approaches. Classical models included Spectral Residual~\cite{hou2007saliency}, Fine Grained~\cite{montabone2010human}, and the Itti-Koch model~\cite{itti1998model}. Deep learning models included TranSalNet~\cite{lou2022transalnet}, a transformer-based model with a ResNet backbone, and UNISAL~\cite{droste2020unified}, a unified image and video saliency model. Each model produced a full-resolution saliency map for every rendered frame across all 256 trials.

Using the object segmentation masks, we extracted per-object saliency distributions at each frame and computed four summary statistics across all frames where the object was visible: mean, median, 95th percentile, and maximum saliency within the masked region. To account for potential confounds arising from object visibility, we measured each object's spatial extent (mean fraction of frame pixels occupied) and temporal extent (fraction of frames in which the object appeared). Saliency scores were residualized against these confounds via ordinary least squares regression to isolate the unique contribution of saliency.

\begin{figure}[t]
    \centering
    \includegraphics[width=\linewidth]{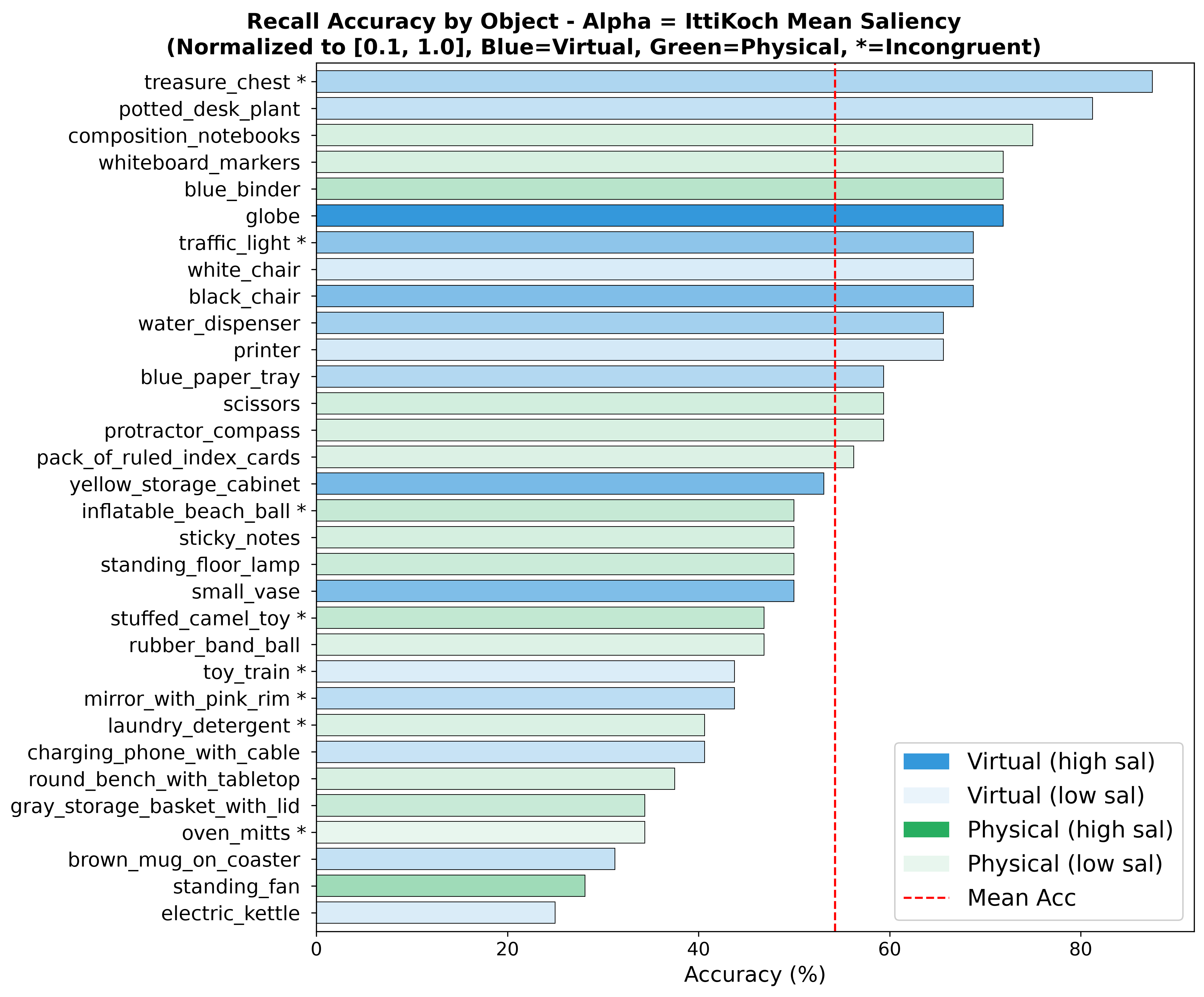}
    \caption{Recall accuracy by object. Bar color indicates virtuality (blue = virtual, green = physical), and bar opacity is proportional to normalized mean Itti-Koch saliency (scaled to [0.1, 1.0]).}
    \label{fig:saliency-recall}
    \vspace*{-3ex}
\end{figure}

A binomial generalized linear mixed-effects model with a logit link was fitted to test whether saliency values from these models predicted the probability of object recall. The model included crossed random intercepts for participants and objects to account for repeated observations involving the same participant and the same object.

Of the five models evaluated, the UNISAL and Itti-Koch models produced the strongest association between computational saliency and object recall, but they were not significant predictors (UNISAL: $\chi^2(1) = 2.11$, $p = .146$, $\eta^2_p$ = .002, marginal $R^2 = .006$, conditional $R^2 = .139$, Itti–Koch: $\chi^2(1) = 1.73$, $p = .188$, $\eta^2_p$ = .001, marginal $R^2 = .006$, conditional $R^2 = .133$) 

This suggests that recall varied more across participants and objects than as a function of any of our salience metrics.



\autoref{fig:saliency-recall} visualizes per-object recall accuracy shown with object virtuality and normalized saliency scores encoded as visual variables.

Overall, saliency only accounts for a small portion of the variance in recall ($r^2 \approx 0.78\%$), consistent with the view that computational saliency captures one component of a multifaceted encoding process shaped by congruence, task demands, and individual strategy.

\subsubsection{Correlation Analysis}
\label{sec:correlation}

We performed a combined analysis of impact factors predicting object recall using our multivariable binomial generalized linear mixed-effects model. The fixed effects comprised object congruence, audio condition, recall cueing, object virtuality, and AR and VR familiarity as described in study design and methodology, normalized exposure time (duration the object was in the center of the view), as well as metrics for object saliency, luminance (perceived brightness relative to its surroundings), and clutter (spatial interference i.e. cognitive effort to distinguish the object from its neighbor) as computed by cv2~\cite{opencv_python} methods. 

Predictor importance was then assessed using semi-partial R². The marginal R² represented the variance explained collectively by the fixed effects, excluding variance attributable to participant and object random effects. A separate semi-partial R² was calculated for each predictor by comparing the complete model with a reduced model from which that predictor was omitted. These values therefore represented each predictor’s unique contribution after controlling for all other predictors.

The complete set of fixed effects explained 6.31\% of the model variance, marginal \(R^2 = .0631\), 95\% CI \([.0548, .1220]\). Normalized exposure time had the largest unique contribution, semi-partial \(R^2 = .0332\), 95\% CI \([.0243, .0935]\), followed by whether the user was alerted about recall, semi-partial \(R^2 = .0091\), 95\% CI \([.0005, .0704]\). The unique contribution of saliency was smaller and uncertain, semi-partial \(R^2 = .0078\), 95\% CI \([.0000, .0693]\). Clutter, \(R^2 = .0040\), luminance, \(R^2 = .0032\), object virtuality, \(R^2 = .0018\), audio condition, \(R^2 = .0010\), VR familiarity, \(R^2 = .0009\), object congruence, \(R^2 = .0001\), and AR familiarity, \(R^2 < .0001\), each made comparatively small unique contributions; all corresponding confidence intervals included zero.

The Type II Wald tests were consistent with exposure time and recall cueing being the clearest predictors. Exposure time was significantly associated with recall, \(\chi^2(1) = 12.44\), \(p < .001\), as was recall cueing, \(\chi^2(1) = 8.38\), \(p = .004\). Saliency showed a positive but inconclusive association, \(\chi^2(1) = 2.92\), \(p = .087\). None of the remaining predictors was statistically significant, all \(p \geq .225\). 

Overall, correlations were weak in magnitude, suggesting that no single factor strongly predicted object recall, with exposure time and recall cueing as the most prominent fixed-effect predictors.


\section{Discussion}

Augmented reality allows for virtual content to be displayed in the real world, and is projected to enable spatial computing. In a digital world, we are faced with a baseline level of attentional impositions that continue to increase. Integrating augmented reality content into the physical world could result in significant consequences for attention.

The main goals of this experiment were therefore to understand the impact of user-level factors, such as task cueing and dual tasking, object-level factors such as size and virtuality, and context-level factors such as congruence on user recall of objects in an attention-demanding AR environment. In order to better understand potential sources of variance not captured by the modeled factors, we conducted a detailed salience analysis after the fact. We were particularly interested in the correlation of salience compared to that of the original modeled factors, and, as the above results show, saliency correlates slightly better with object recall than object virtuality, object size, presence of audio task, and participant AR familiarity. The correlation is slightly lower than the one for exposure time and task cueing. Overall, the $R^2$ values are fairly small, which indicates that for our experiment setup no single factor, not even natural object salience, is a great predictor for participant object recall. However, we believe that strongly salient explicit AR highlighting methods (such as animated high-contrast alerts) will very likely modify the underlying recall distribution resulting in higher correlated recall. One outcome of this study is in fact the realization that too subtle highlighting methods might not move the needle on recall much, so stronger, more pronounced highlighting should be tried.

A main motivation for this work was to explore the impact of object virtuality on user attention (as measured indirectly through object recall) in an environment densely populated  with physical and virtual objects. The results indicate a main effect of virtuality, which is explained by significant interactions between virtuality and both object size and semantic congruence. Specifically, virtual objects were recalled more accurately than physical objects when the objects were medium or large in size, and this advantage was also observed for semantically incongruent objects. These findings are consistent with prior literature suggesting that, under certain conditions, virtual objects attract greater attention than physical objects. That said, our correlation analysis for impact factors on object recall (see Section \ref{sec:correlation}) lists virtuality as having insignificant impact.

We observed improved performance on recall of objects present in the environment when participants were cued to the memory task as compared to trials when they were not cued, which aligned with our hypotheses. Surprisingly, the performance gains in the cued trials were lower than might have been reasonably expected given that participants were specifically instructed to pay attention to the environment. This points to the presence of external and intrinsic factors that minimized users' ability to commit objects in the environment to memory despite explicit verbal instructions.

We observed reduced performance in the primary selection task when the simultaneous audio task was included, validating our choice of the dual task for increasing attention demand. Additional cognitive load was likely induced by the high density of objects to be attended to within a short trial duration. 

Given these attentional demands, participants would have needed to prioritize the three tasks (selection, audio and recall) in a sequence to be able to complete any of them successfully. We did not prime participants to the importance of the three tasks, letting them determine their own priorities based on multitasking ability and intrinsic motivation. Our findings indicate that participants who prioritized the memory task over one or more of the other tasks performed better at that task in the cued trials compared to participants who gave it the least priority, demonstrating that task cueing, though effective, is only one factor that influences attention in a multitasking scenario. 

Intrinsic and reward-based motivation has been shown to influence prioritization of items in working memory tasks~\cite{grogan2022motivation, atkinson2024does}, but intrinsic motivation is highly dependent on individual factors that may not be easily influenced by instructional cues. It is therefore important to consider the use of visual cues which can support bottom-up attention capture in conjunction with instructional cues to direct attention in cognitively demanding environments.

From a design perspective, these results carry important implications for AR interface development. Attention related cues, including visual and semantic cues, can provide support for users where encoding and recalling environmental details are involved. However, passive use of AR interfaces without intentional guidance may result in impacted attention to the backdrop environment resulting in low object recall especially for large but congruent physical items that blend into the background (such as the floor lamp and standing fan which over half of the participants did not recall correctly). This underscores the need for AR systems to actively direct user attention toward mission-critical scene elements rather than relying on visual presence alone.



In this study, object congruence emerged as a significant factor in influencing recall but only for physical objects. Congruent physical objects were remembered more accurately than incongruent ones. Confidence ratings are in sync with our findings: participants reported higher confidence for congruent present objects. This underscores how congruence can modulate attention via top-down expectations, potentially leading participants to feel more certain about familiar or congruent items.
Our congruency findings align with the literature around the consistency effect. Schemas support memory for expected objects by providing retrieval structure and enabling reconstruction \cite{brewer1981role, webb2020memory}. On the other hand, literature around attentional capture theories reports that semantically unexpected objects stand out and are preferentially encoded \cite{chen2019attentional}. In this context, our reported interaction between virtuality and congruence is of interest since it may indicate situational preference of either mechanism (consistency effect and schema incongruity): With physical objects, we saw a more accurate recall of congruent objects, whereas with virtual objects that was not the case (see also Figure \ref{fig:cue}, Left).

\section{Limitations}

While we purposefully created an attention-demanding task and environment, by placing many objects in a small office space, introducing three parallel tasks (object spotting and selection, audio call-sign response, and object recall), providing time pressure by limiting trial length to 45 seconds, asking about a large number of objects and thus exceeding short-term memory limits, we unfortunately couldn't exert sufficient experimental control to be able to tease apart the exact contributions of each of these design decisions. Systematically varying the presence of the call-sign task and administering questionnaires about participants' prioritization among the parallel tasks at least gave us some experimental control.  

There may also have been a retroactive interference effect, combined with attentional resource competition: Later task demands (e.g., object recall) may have been impaired because earlier encoding of another large set (object selection/search) occupies working memory and disrupts consolidation. It would be interesting to conduct a similar study with a different foreground task that does not enumerate objects in the environment.

While our design choices reflect real-world multitasking demands, they may have limited participants’ ability to prioritize recall, hence providing an explanation for why we do not see as large of an impact of cueing as we had originally expected.

Another design limitation was the cueing condition order. The study relied on a surprise recall paradigm, so the order of conditions had to remain fixed (no cue followed by cue). This design ensured the validity of the cueing effect to remain within subjects but limited our ability to counterbalance order and tease apart the impact of possible practice effects and cueing. However, we can basically rule out large practice effects on the object recall task: we didn't see a large effect size for cueing (the later condition) and in general, participants' object recall wasn't better the second time around. Also, as a reminder, the memory recall assessment occurred twice on disjoint sets of objects and was deemed difficult by the participants. 

We would like to have more direct evidence for what aspects of our findings are specifically AR-related. To this end, we would like to run a similar study without AR technology involved, which of course would complicate the physical shuffling and task setup drastically. Similarly, we would like to run a comparison study in which we use AR for highlighting certain object recall candidates (physical or virtual) during the foreground task and see if their recall rates can be boosted by this AR intervention.

\section{Conclusion}

This study demonstrated that attention to objects in a dense office-themed environment is influenced by object-, user- and context-level features. 
Virtual objects were recalled more accurately than physical objects, particularly for medium and large objects. Among virtual objects, larger items were also recalled better than smaller ones, although object size alone did not produce a significant main effect.
Congruent items were better encoded for physical objects compared to virtual ones,
which emphasizes the influence of directed attention in an AR interface. Cueing improved memory for present objects suggesting that cues reinforced attention to objects within the scene, and that our study design was reasonably attention-demanding with a high density of visual information. We also found weak positive correlation of object saliency with recall, although the association did not reach statistical significance. These findings have implications for AR interface design: systems should incorporate attention-informed elements (if recall is paramount, a high salience boost may be necessary) while minimizing cognitive distractions to support user memory and task efficiency.

To probe cognitive load more rigorously, future iterations of the study could increase task difficulty, potentially leveraging a controllably hard task such as an N-back task, and examine effects across various levels of cognitive load. Investigating both implicit (non-verbal) and explicit (verbal) cueing can reveal how different attention strategies can impact encoding. Additionally, controlling for potential learning effects across repeated selection trials will improve interpretation of performance trends. Incorporating physical walking during the task could serve as an added task load in a future study, which could help determine how movement affects memory and object interaction in AR contexts. Future research can further refine our understanding of human cognitive processes in AR and inform the design of adaptive, cognitively informed AR systems.

\acknowledgments{%
This work was funded in part by the DARPA Intrinsic Cognitive
Security Program under grant number \#HR001124C0409, via subcontract from SRI International, as well as ONR award N00014-23-1-2118 and NSF award IIS-2211784.
}

\bibliographystyle{abbrv-doi-hyperref}

\bibliography{references}

\end{document}